# Mapping the Digital Diplomatic Infrastructure:
## A Comparative Evaluation of
## Global Online Directories for Diplomatic Missions

Siniša Grgić, Ph.D.[1]
Ambassador of the Republic of Croatia

*This study provides a comparative evaluation of global diplomatic mission directories. DiplomaticMonitor.org, EmbassyPages.com, and WikiData.org are strategically selected among the top ten global services. After analyzing nearly all available online global diplomatic directory services, these three platforms are selected as they represent fundamentally different approaches to creating worldwide diplomatic mission databases. Using official diplomatic lists from over 150 countries as benchmarks, we assessed data coverage, accuracy, and update frequency across these platforms. DiplomaticMonitor consistently outperforms its counterparts in structure, completeness, and timeliness, accurately reflecting ambassadorial appointment cycles and maintaining high precision across contact and personnel records. EmbassyPages, despite strong search engine visibility and widespread usage, exhibits significant data currency issues, with markedly diminished ambassadorial accuracy attributable to delayed refresh cycles. WikiData offers valuable historical documentation and open-source accessibility but lacks the consistency and verification protocols necessary for reliable real-time diplomatic information. Our findings highlight the critical challenge posed by the absence of a standardized global diplomatic mission registry. In this fragmented landscape, methodologically rigorous third-party platforms can occasionally surpass government-published records in quality and utility. The research demonstrates that in contemporary digital diplomacy, data reliability correlates less with institutional provenance than with disciplined, transparent, and consistent data stewardship practices.*

## 1. Introduction

In an era of accelerating digital transformation, accurate and accessible information about diplomatic missions has become a vital component of international engagement. From foreign ministries and consular services to researchers, journalists, and citizens abroad, a broad range of users increasingly rely on digital platforms to locate embassies, consulates, and diplomatic personnel around the world.

Yet, from the author's extensive experience in foreign service, there has been a consistently encountered challenge: online users often find unofficial or outdated listings on third-party websites before locating our official embassy websites. This confusion is primarily due to the inconsistent digital visibility of government platforms, which frequently change domains, restructure ministries, or shift institutional responsibilities without prioritizing search engine optimization (SEO) or global discoverability.

---

[1] more about the author: sinisagrgic.net

By contrast, many private directory services have gradually built strong brands and continuously improve the quality of their content. Because the underlying data is public, it is not uncommon for some platforms to copy large portions of content from more credible and diligently maintained sources and present it as their own. This practice results in fragmentation, and worse, leads to outdated records as many of these services are poorly maintained after initial data harvesting.

One of the most pressing issues is data decay. Diplomatic personnel, particularly ambassadors, are appointed for relatively short mandates – on average four years – compared to CEOs of major corporations, whose average tenure can extend to 7.2 years for S&P 500 companies, with top performers serving an average of 15 years (Citrin, Hildebrand, & Stark, 2019). This stark difference in tenure stability means that more than 25% of ambassadorial data becomes outdated within just one year, and diplomatic listings degrade at a rate of over 2% per month if not actively maintained.

Adding to the complexity is the absence of a single global "official register" of diplomatic missions. Each country maintains its own records, often in different formats, languages, and levels of completeness. Some publish only addresses, others include phone numbers or emails, while some omit contact details entirely. Even the formatting varies – from fully functional websites to PDFs, Word or Excel documents, and in some cases, completely unstructured text. Diplomatic lists might include the spouse's name, date of accreditation, or present surnames in uppercase, but practices vary widely.

Because of all these factors, third-party online directories – if managed carefully and regularly updated – can play an essential role in supporting the global diplomatic ecosystem. This paper seeks to critically evaluate ten of the most prominent platforms offering such services and to identify their comparative strengths, limitations, and unique features using a robust set of methodological metrics.

This study builds upon a rich tradition of diplomatic data collection and analysis, particularly the pioneering work done by the Frederick S. Pardee Center for International Futures at the University of Denver. Their Diplometrics project (Moyer, Bohl, & Turner, 2015) represented a significant milestone in systematically documenting diplomatic representation across time, updating and extending the foundational efforts of Singer and Small (1966, 1973) and Bayer (2006) through the Correlates of War Project. The Diplometrics dataset tracked annual diplomatic relationships between 1960 and 2013, offering unprecedented granularity by distinguishing levels of representation, from ambassadors to chargés d'affaires, and capturing the physical presence of missions. This meticulous approach to diplomatic data laid important groundwork for understanding global diplomatic networks, though it primarily focused on official bilateral representation rather than the digital infrastructure that now mediates access to this information.

Other scholarly work has further expanded our understanding of diplomatic networks through various analytical approaches. Kinne's (2014) examination of diplomatic signaling and prestige dynamics complements the structural data foundation, while Hafner-Burton, Kahler, and Montgomery (2009) pioneered network analysis methodologies that have transformed how we understand these diplomatic connections. These approaches help contextualize our current investigation into the quality of digital diplomatic information available to the public through third-party platforms, which represents a natural evolution from documenting the networks themselves to examining how they are represented in the digital sphere.

## 2. Methodology

This study adopts a comparative approach to assess the structure, coverage, and accuracy of three widely used online services offering global data on diplomatic missions: DiplomaticMonitor, EmbassyPages, and WikiData (Wikipedia's dataset of diplomatic missions). These platforms were selected based on their global prominence, distinct data models, and relevance to both public and professional users. By focusing on this trio, the study enables a detailed, high-resolution comparison while maintaining analytical clarity.

The evaluation was conducted across two primary dimensions. The first dimension focused on data coverage, assessing the quantitative breadth and depth of each platform's information. This included the total number of countries covered, as well as the number of embassies, consulates-general, consulates, and other diplomatic or multilateral missions listed. Further analysis was conducted to determine the number of missions represented on each continent and the volume of contact information available – such as physical addresses, telephone numbers, email addresses, and official websites. In addition, the study measured how many individual diplomats were listed per platform, distinguishing between heads of mission (ambassadors or equivalents) and other diplomatic personnel, thus offering a view of not only structural data but also personnel-level detail.

The second dimension evaluated data quality, particularly the accuracy and timeliness of each platform's content. To establish a baseline for comparison, the research team compiled a reference dataset based on the official diplomatic lists published by foreign ministries and protocol departments in more than 150 countries. These lists were obtained from government websites, including web pages, downloadable documents (PDF, Word, Excel), and structured databases, wherever available. These were then compared to entries on the selected third-party platforms in order to detect discrepancies, identify delays in updates, and evaluate structural consistency.

A key aspect of this analysis involved the detection of changes in ambassadorial appointments and mission statuses over time. To support this work, the author engaged two professional data-entry teams through the UpWork platform: Navin Mirania and his team from India, and Shahid Zaman and his team from Pakistan. These teams were tasked with entering and organizing data from hundreds of verified news sources, foreign ministry announcements, and press releases documenting the presentation of credentials by ambassadors. Their work did not involve any analytical or interpretive elements but was essential in ensuring that the raw data required for accuracy testing was thorough and well-structured.

Data collection and parsing involved a combination of tools and methods, including manual inspection, spreadsheet-based tabulation, and where permitted, automated data extraction techniques. Attention was paid to standardize country names, mission types, and personnel titles to ensure comparability across the platforms. Ambiguities in mission naming, address formatting, or accreditation details were resolved by deferring to the most recent government-published lists as the authoritative source. To address the challenge of identity resolution across disparate data sources, we employed the Jaro-Winkler string similarity algorithm to compare heads of missions' names (Cohen, Ravikumar, & Fienberg, 2003). Name pairs yielding similarity scores above 0.85 were classified as identical entities, while those below this threshold were treated as distinct individuals. Validation testing indicated that this approach maintained an error rate below 1%, which had negligible impact on our comparative analysis while significantly reducing manual verification requirements.

The methodology used here is consistent with open data benchmarking practices and international digital standards, including principles drawn from the Open Data Barometer and the FAIR data framework. Although constrained by the lack of access to platform-specific update logs or internal APIs, the study compensates for this through triangulation with official records and manual verification.

While recognizing the inherent variability in the publication practices of foreign ministries worldwide – ranging from structured, multilingual portals to non-standardized, monolingual PDF lists – the methodology aims to offer a fair and replicable comparison of platforms that claim to consolidate this information for global users. The results, presented in the following sections, provide insights into both the current state and the future potential of digital diplomatic infrastructure.

### 3. Overview of Evaluated Platforms

This study conducted a comprehensive evaluation of ten prominent online platforms providing diplomatic mission information globally. These platforms represent a diverse ecosystem of services catering to different user needs – from travelers seeking consular assistance to researchers documenting diplomatic networks. Each platform was assessed for content coverage, update frequency, data structure, and user interface design to establish a baseline understanding of the current digital diplomatic information landscape.

**DiplomaticMonitor.org** is a professionally managed platform that stands out for its country-to-country architecture and near real-time updates. Inheriting the legacy of the renowned Diplomacy Monitor, the site aggregates and validates data by scraping official diplomatic lists directly from more than 180 countries' protocol departments. It offers detailed embassy and consulate records, including contact information and accredited personnel, and is distinguished by its data transparency and responsiveness. Among currently active platforms, it is arguably the most accurate and technically robust.

**Embassies.net** is a content-rich directory that includes embassies, consulates, and visa-related information across most countries and territories. Its user interface is well-organized and easy to navigate, offering contact details, accreditation info, and links to official mission websites. It is one of the few directories that attempts to separate resident and non-resident missions clearly. While it does not disclose its data sources or update schedules, the site appears to be actively maintained and is frequently referenced in public search results.

**Embassy-Finder.com** offers a minimalist and fast-loading interface that provides basic information about embassies and consulates by country. The site emphasizes simplicity and quick look-up features but lacks structural depth and does not include diplomatic personnel data. It has limited integration with official sources and rarely includes full contact details beyond addresses and phone numbers. It serves primarily as a first-contact tool but is inadequate for in-depth research or official referencing.

**EmbassyPages.com** is one of the most widely used private directories of embassies and consulates worldwide. The platform offers clear listings by country and city, including contact details, accreditation status, and links to mission websites. Although not affiliated with any government, it has built a strong digital reputation through SEO optimization and a consistent user experience. However, it does not cite official sources and offers limited transparency about update cycles, which can affect data reliability over time.

**EmbassyWorld.com** serves as a basic global directory of diplomatic missions organized by country. The platform offers fundamental contact details and addresses but suffers from significant updating deficiencies. Its simplistic interface prioritizes basic functionality, yet many listings contain severely outdated information, including obsolete contact details dating back decades. Despite reasonable coverage of major diplomatic posts, its practical utility is compromised by these persistent data currency issues.

**Embassy-Worldwide.com** is a searchable database of embassies and consulates, organized by host and sending countries. While it covers a wide range of missions and includes essential contact details, the platform design is dated and offers minimal contextual information. It is useful for basic lookup purposes but lacks advanced features or detailed accreditation metadata. Update frequency is not disclosed, and the accuracy of the listings varies by region.

**Foreign-Trade.com/Resources/Embassy.htm** is a static directory embedded within a broader trade-oriented website. It focuses primarily on embassies in Washington, D.C., and U.S. missions abroad, reflecting its primary audience of international business users. While the list is practical for quick reference, it is limited in geographic scope and depth. The platform does not offer interactivity or structured data tools, and the listings are not regularly verified against official updates.

**VisaHQ.com** is primarily a visa processing service that includes embassy and consulate information as part of its application workflow. While not designed as a dedicated diplomatic directory, its listings are accurate for countries with high visa demand, particularly for travel to or from the United States, the UK, and Schengen Area nations. The platform integrates embassy addresses and sometimes personnel contacts, but coverage is uneven globally and reflects the company's commercial priorities.

**WikiData.org (Wikipedia)** provides an expansive, crowd-sourced collection of embassy and consulate data, organized by both sending and host country. It is the most comprehensive open-access resource in terms of geographic and historical coverage, and frequently includes additional contextual notes, photos, and links. However, the decentralized nature of editing leads to inconsistencies in format, data quality, and update frequency. While invaluable as a general reference, it is not formally curated and should be verified against official sources for precision.

**Xpatulator.com** is a relocation support platform that includes diplomatic mission data as part of its country profiles. The site is designed for expatriates and corporate relocation planners, offering embassy and consulate listings as supplementary content alongside cost-of-living, housing, and tax information. It is not a specialized diplomatic tool, and its embassy data is neither comprehensive nor regularly updated, but it can provide a useful entry point for users planning moves abroad.

While all ten platforms were thoroughly analyzed using consistent methodological criteria, this paper focuses on a comparative evaluation of three representative services: DiplomaticMonitor, EmbassyPages, and WikiData. These platforms were selected based on their distinct operational models, differing data collection methodologies, and varied user bases, providing an optimal cross-section of the digital diplomatic directory ecosystem.

DiplomaticMonitor represents newer, technology-driven approaches with structured data harvesting; EmbassyPages exemplifies established commercial services with strong search engine visibility; and WikiData offers insights into crowdsourced, open-data models. By concentrating our detailed analysis on these three distinct approaches, we can more effectively identify patterns and principles that apply across the broader landscape of diplomatic directory services.

**4. Comparative Analysis**

To evaluate the relative performance of the three selected platforms – DiplomaticMonitor (DM), EmbassyPages (EP), and WikiData (WD) – two comparative tables were developed: one measuring data coverage and the other assessing data quality. These tables, built from structured extraction and manual validation against official sources, reveal significant differences in scale, accuracy, and structural depth among the platforms.

*Data Coverage*

The first table illustrates the overall volume and structure of the data provided by each platform. DiplomaticMonitor leads across nearly all indicators, with over 34,000 total missions, including embassies, consulates-general, consulates, and honorary consuls. EmbassyPages ranks second with approximately 28,000 missions, while WikiData lags behind with only 10,457, reflecting its community-driven model that predominantly captures embassies while underrepresenting other missions such as honorary consulates and standard consular offices.

**Table 1.** Data Coverage. *(DM = DiplomaticMonitor; EP = EmbassyPages; WD = WikiData)*

|  | DM | EP | WD | EP/DM | WD/DM |
|---|---:|---:|---:|---:|---:|
| Missions | 34,374 | 28,379 | 10,457 | 82.6% | 30.4% |
| Non-Residential Coverage | 6,417 | 3,625 | 6,693 | 56.5% | 104.3% |
| Bilateral Relations | 17,437 | 15,414 | 8,893 | 88.4% | 51.0% |
| Countries & Territories | 218 | 218 | 205 | 100.0% | 94.0% |
| Distinct Places | 3,737 | 1,861 | 840 | 49.8% | 22.5% |
| Embassies | 12,237 | 10,926 | 8,757 | 89.3% | 71.6% |
| Consulates-General & Consulates | 4,669 | 3,584 | 1,186 | 76.8% | 25.4% |
| Honorary Consuls | 17,055 | 13,162 | 272 | 77.2% | 1.6% |
| Missions in Australia & Oceania | 725 | 580 | 238 | 80.0% | 32.8% |
| Missions in North America | 3,018 | 2,465 | 891 | 81.7% | 29.5% |
| Missions in South America | 2,466 | 1,949 | 691 | 79.0% | 28.0% |
| Missions in Europe | 15,271 | 12,450 | 4,074 | 81.5% | 26.7% |
| Missions in Africa | 5,228 | 4,474 | 1,577 | 85.6% | 30.2% |
| Missions in Asia | 7,229 | 6,170 | 2,918 | 85.4% | 40.4% |
| Missions with Telephone(s) | 30,662 | 26,150 | 983 | 85.3% | 3.2% |
| Missions with Email(s) | 30,221 | 23,539 | 1,048 | 77.9% | 3.5% |
| Missions with Website(s) | 11,957 | 6,031 | 6,700 | 50.4% | 56.0% |

In terms of non-residential coverage – missions accredited from third countries – DiplomaticMonitor again dominates with 6,417 records, compared to 3,625 for EmbassyPages. WikiData interestingly exceeds both in this category with 6,693, suggesting some contributors may be entering historical or broader-accreditation records without uniform verification standards. However, WikiData's practical utility is significantly diminished by outdated jurisdictional information and inconsistent formatting, with entries often cluttered by unstructured local and regional jurisdictional descriptions that compromise data usability and reliability.

Coverage of bilateral diplomatic relations reinforces DiplomaticMonitor's lead, listing 17,437 active bilateral entries, while EmbassyPages provides 15,414 and WikiData only 8,893, or about 51% of the DM total. All three platforms reflect comprehensive country and territory coverage (205–218 entries), but diverge widely in the number of distinct cities and locations where missions operate: DiplomaticMonitor lists 3,737, EP only 1,861, and WD just 840. When categorized by mission type, the disparities become more evident. DiplomaticMonitor includes over 12,200 embassies, 4,669 consular missions, and a notably large pool of 17,055 honorary consuls – more than four times the count available in WikiData. EmbassyPages follows closely in embassies and consulates, but significantly underrepresents honorary consuls. WikiData's low count of 272 honorary consuls reveals its limited scope in capturing the broader diplomatic ecosystem. Regional coverage reinforces these trends: across all six continents, DiplomaticMonitor consistently records between 14% and 20% more missions than EmbassyPages and 2-3 times more than WikiData. This includes over 15,000 missions in Europe, more than 7,000 in Asia, and substantial entries in Africa, the Americas, and Oceania.

In terms of contact details, the discrepancies are stark. DiplomaticMonitor offers telephone data for over 30,000 missions, email addresses for 30,221, and websites for nearly 12,000 missions. EmbassyPages, while strong in telephone data (26,150) and emails (23,539), trails significantly in websites (6,031). WikiData performs weakest in these operational categories, with fewer than 1,100 entries for either emails or telephone numbers, although its website count (6,700) slightly exceeds that of EmbassyPages, likely due to crowd-sourced linking to official domains.

*Data Quality*

The assessment of data quality centers on two core indicators: the recency of diplomatic appointments and the accuracy of contact details, both compared to a verified reference dataset derived from official diplomatic lists collected in December 2024. One of the most telling variables is the median accreditation year for listed diplomats, which serves as a proxy for how current the data is on each platform.

**Table 2.** Data Quality. *(DM = DiplomaticMonitor; EP = EmbassyPages; WD = WikiData)*

|  | DM | EP | WD |
|---|---|---|---|
| Median Accreditation Year | 2021 | 2019 | 2017 |
| Median Accred. Year for Ambassadors | 2022 | 2020 | 2017 |
| Matching HoMs | 98% | 84% | n/a |
| Matching Ambassadors | 98% | 38% | n/a |
| Matching Telephones | 99% | 95% | n/a |
| Matching Emails | 94% | 86% | n/a |
| Matching Websites | 87% | 84% | n/a |

In this regard, as presented in the second table, DiplomaticMonitor again leads, with a median accreditation year of 2021 for all heads of mission (HoMs) and 2022 for ambassadors. EmbassyPages shows slightly older entries (2019 and 2020, respectively), while WikiData reflects a static median year of 2017 across the board. These values reveal not only the frequency of updates but also the internal composition of each dataset. For instance, the median accreditation year for HoMs includes honorary consuls, many of whom are appointed for life or serve indefinite terms. These lifetime appointments tend to lower the average, yet the gap between the HoMs' and ambassadors' median years remains relatively narrow – suggesting that many countries do not include honorary consul accreditation dates in their public records, or that they are underreported in the datasets.

Chart 1 provides compelling empirical evidence of the timeliness differential between platforms. Given the standard four-year diplomatic cycle and this study's 2024 data collection endpoint, the expected benchmark for current ambassadorial accreditation centers around 2022. DiplomaticMonitor demonstrates exceptional currency, closely tracking this natural turnover pattern. The visualization quantitatively demonstrates a consistent two-year lag in EmbassyPages's data refresh cycle compared to DiplomaticMonitor.org when analyzing identical diplomatic representatives across both platforms. This temporal disparity manifests in significant accuracy degradation – only 38% of ambassadors listed in EmbassyPages corresponded with official appointments, whereas DiplomaticMonitor achieved 98% correspondence. With each year of delay rendering approximately 25% of ambassadorial records obsolete due to the turnover cycle, this two-year lag mathematically explains the substantial precision differential between these platforms. The identification of identical representatives was methodologically sound, as accreditation dates from DiplomaticMonitor were used for temporal placement of matched individuals, given EmbassyPages's omission of appointment chronology.

**Chart 1.** Timeliness of platforms' content. *(Number of Heads of Mission by accreditation year)*

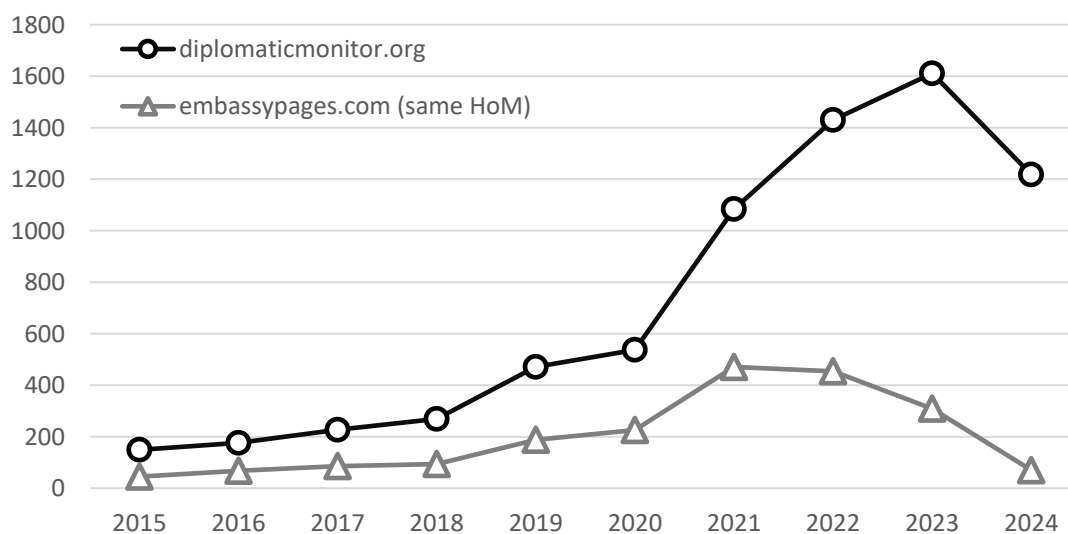

WikiData's persistent median year of 2017 is largely due to the low number of honorary consuls in its dataset – just 272 entries, compared to over 17,000 in DiplomaticMonitor and 13,000 in EmbassyPages. The relatively low presence of consular records, combined with the community-driven nature of the platform, limits its ability to keep pace with fast-changing diplomatic appointments.

It is important to acknowledge that accuracy comparisons with government-published diplomatic lists have inherent limitations. While official lists are considered authoritative, they are often released with delays, published in various formats, or lack standardization. In some cases, third-party platforms – especially those actively managed and in regular contact with diplomatic missions – may actually maintain fresher or more granular data than certain foreign ministries. Therefore, while this study treats official sources as the benchmark, it also recognizes that private platforms may surpass them in operational utility if they commit to continuous updates.

When examining contact details – such as telephone numbers, emails, and websites – the variance in accuracy is less dramatic. These types of data tend to change less frequently than ambassadorial appointments, and their quality depends more on the effort spent in aggregating and validating information. DiplomaticMonitor leads in all three categories, with 99% accuracy for phone numbers, 94% for emails, and 87% for websites, all validated against a 15,000-record sample. EmbassyPages also performs reasonably well, especially for telephone numbers (95% match rate), though it also has high accuracy for telephones and emails. The modest differences in this domain suggest that both platforms use persistent and (semi-)automated collection methods, though the gap widens in areas requiring sustained manual upkeep or direct mission engagement.

*Interpretation*

The comparative findings presented above highlight profound differences not only in the scale of diplomatic data aggregation but also in the philosophies, operational models, and long-term sustainability of each platform. DiplomaticMonitor clearly emerges as the most structured, comprehensive, and accurate among the three analyzed services, setting a new benchmark for what a digital diplomatic infrastructure can achieve when rooted in systematic updates and official-source integration.

Its extensive reach – covering more than 34,000 diplomatic missions, including consulates, embassies, and over 17,000 honorary consuls – is unmatched in both volume and granularity. This level of completeness is not just the result of broad scraping, but of a carefully maintained and frequently validated system that tracks accreditation data in near real time. The fact that DiplomaticMonitor's median ambassador accreditation year is 2022, precisely matching the expected benchmark given typical four-year diplomatic terms and a data collection cut-off in December 2024, reinforces its credibility.

By contrast, EmbassyPages, while widely recognized and long-established, suffers from a slower update cycle and a limited feedback mechanism with official sources. The platform's two-year lag in ambassadorial updates is not a minor discrepancy – it directly impacts data utility, especially for users who depend on the accuracy of heads of mission listings for diplomatic correspondence, protocol events, or academic research. The drop in ambassadorial accuracy to 38% suggests that unless significant effort is invested in data maintenance, the natural diplomatic turnover rate will continue to degrade the platform's reliability. This also illustrates a broader truth: in the world of diplomacy, where ambassadorial churn is both expected and high, even a one-year delay can result in a quarter of the records being outdated.

WikiData, while the most open and participatory model among the three, reveals the limitations of crowd-sourced data for high-precision domains such as diplomacy. The platform's strong historical and encyclopedic value cannot be overstated – it includes rare or historic missions and offers useful links and context – but its capacity to reflect current global diplomatic structures is inherently constrained. With only 272 honorary consuls listed and outdated ambassadorial data (median year 2017), WikiData lacks the infrastructure and incentives needed to support regular updates on thousands of dynamic, bilateral relationships.

Interestingly, while discrepancies in personnel data are stark, the contact information (emails, phones, and websites) tells a different story. These details tend to be more stable over time and less influenced by political turnover. EmbassyPages achieves moderate success in this area, particularly for mission websites, while DiplomaticMonitor consistently ranks highest across all three fields. This reinforces a secondary but important finding: platforms that invest in broader sourcing – including direct outreach to missions, integration of local directories, or manual verification from semi-official sources such as yellow pages – can provide longer-lasting utility, even in less-structured data environments.

Finally, the study acknowledges that no single benchmark – including official diplomatic lists – is flawless. In practice, some government-published lists are updated infrequently, vary greatly in formatting, and may omit key data points like honorary consul appointments or exact accreditation dates. It is therefore plausible – and even likely – that a well-maintained third-party platform can outperform official sources in certain areas, especially if it fosters ongoing relationships with embassies, monitors government communications continuously, and applies consistent editorial standards.

The findings underline the importance of data stewardship over mere data aggregation. The difference between an outdated diplomatic directory and a reliable one is not only in initial coverage but in the invisible, ongoing labor of revision, verification, and responsiveness to geopolitical change. DiplomaticMonitor exemplifies this proactive model, while EmbassyPages and WikiData represent legacy and communal approaches that may still serve important roles – but increasingly require modernization to remain relevant in a rapidly shifting diplomatic landscape.

## 5. Limitations & Considerations

While this study offers a high-resolution comparison of three leading platforms for global diplomatic mission data, several limitations – both structural and methodological – must be acknowledged to contextualize its findings and guide future research.

First, the absence of a unified, official global registry of diplomatic missions imposes inherent constraints on any comparative effort. Each country maintains its own diplomatic list, often in non-standardized formats and in multiple languages, with varying degrees of completeness, update frequency, and public accessibility. Some lists are published monthly in structured formats; others appear sporadically as PDF scans, Word documents, or embedded text on government websites. In many cases, crucial data such as accreditation dates, contact emails, or names of honorary consuls are either missing or inconsistently published. This fragmented landscape complicates the establishment of a fully authoritative baseline for benchmarking.

Second, while the study relied on official diplomatic lists to evaluate accuracy and freshness, these sources are not without their own flaws. Some ministries publish outdated information, delay updates due to political transitions, or exclude non-resident and honorary representations altogether. Consequently, while treated as the benchmark for comparison, these official records are not immune to error, and in specific instances, third-party platforms may surpass them in operational reliability – particularly when those platforms are in regular contact with missions and actively manage corrections based on direct feedback.

A further limitation lies in the temporal nature of the data collection process. All platform evaluations were based on information available in December 2024, offering a snapshot in time rather than a longitudinal analysis. Given the dynamic nature of diplomatic postings – where ambassadorial and consular appointments can change monthly – the conclusions drawn here may shift significantly within a matter of weeks. Future studies would benefit from a time-series approach to track accuracy decay and update responsiveness over longer intervals.

The study also relied on external data-entry teams to support the structuring of raw input from government sources. While these teams were carefully instructed and quality-checked, manual processes always carry the risk of transcription errors, especially when working across multiple languages and document types. To mitigate this, multiple rounds of validation were conducted, but the possibility of residual inconsistencies cannot be entirely ruled out.

In terms of platform-specific constraints, none of the three evaluated services offers transparent APIs, changelogs, or public editorial workflows, making it difficult to assess how frequently updates are made or whether corrections are retroactively applied. For example, while DiplomaticMonitor demonstrates high accuracy and coverage, its internal processes are proprietary and therefore not independently auditable. Similarly, EmbassyPages does not disclose its data collection methods, and WikiData, while open by design, suffers from a lack of systemic oversight and standardization.

This study comprehensively examined all diplomatic mission types, including bilateral embassies, consular offices, and permanent missions to international organizations (U.N., E.U., A.U., etc.). Our analysis encompasses all 193 UN member states and several partially recognized countries and territories that maintain diplomatic relations with certain states. This inclusive approach provides a complete picture of how different platforms handle the full spectrum of diplomatic representation, including complex cases outside traditional state-to-state relationships.

In light of these limitations, the findings of this study should be interpreted not as a definitive ranking of platforms, but as a structured contribution toward understanding the digital maturity and operational integrity of diplomatic data services. As digital diplomacy evolves, maintaining high-quality, regularly updated, and publicly accessible directories will be essential – not just for governments and researchers, but for the global public who increasingly rely on these platforms for trusted information.

## 6. Conclusion

This study set out to evaluate the reliability, scope, and operational integrity of three major global platforms offering structured information on diplomatic missions: DiplomaticMonitor, EmbassyPages, and WikiData. By analyzing both the volume of available data and its accuracy when compared with official diplomatic lists, the research has demonstrated significant variation in the capacity of these platforms to meet the evolving needs of the global diplomatic ecosystem.

The results reveal a clear distinction between platforms that are actively curated and systematically updated and those that rely on passive or community-driven models. DiplomaticMonitor emerged as the most comprehensive and accurate of the three, reflecting a data strategy anchored in institutional diligence, automated monitoring, and synchronization with national diplomatic sources. EmbassyPages, though widely used and intuitively organized, showed signs of data aging, particularly in the accuracy of ambassadorial appointments. WikiData, while commendable for its open-access ethos and historical scope, proved inadequate for real-time diplomatic referencing, largely due to the lack of editorial oversight and structural consistency.

One of the study's most significant findings is that the value of a diplomatic directory lies not simply in its breadth, but in the frequency and quality of its updates. The high turnover of diplomatic personnel – especially ambassadors – renders even moderately delayed datasets obsolete. A one-year delay can result in 25% of records becoming outdated, with direct consequences for diplomatic correspondence, consular coordination, and public trust in digital reference tools.

Moreover, the absence of a single, standardized, global registry of diplomatic missions – paired with inconsistencies in how governments publish their own diplomatic lists – leaves a vacuum that private platforms have attempted to fill with varying degrees of success. The study also underscores a counterintuitive but crucial insight: a well-managed third-party directory, if based on disciplined methodology and constant engagement with official sources, can outperform some government-published lists in terms of usability and accuracy.

As diplomacy becomes increasingly digitized, the tools that support it must evolve in parallel. Future research should continue to monitor the comparative performance of these platforms, while encouraging the development of open, verifiable, and interoperable diplomatic data systems. Ideally, foreign ministries and international organizations should begin to view these platforms not as peripheral services but as potential partners in global diplomatic transparency and public communication.

The findings here are intended to inform practitioners, policymakers, researchers, and developers alike. In doing so, the paper contributes to a growing recognition that in the age of digital diplomacy, data stewardship is no longer a back-office function – it is a front-line responsibility.